\documentclass[aps,prl,twocolumn,showpacs,superscriptaddress]{revtex4}
\usepackage{graphicx}
\bibstyle{apsrev.bib}

\newcommand{\WIDTHAB}{0.8\textwidth}
\newcommand{\WIDTHB}{0.48\textwidth}

\begin{document}
\title{RNA denaturation: excluded volume, pseudoknots and
transition scenarios}
\author{M.~Baiesi}
\affiliation{INFM-Dipartimento di Fisica, Universit\`a di Padova,
I-35131 Padova, Italy.}
\author{E.~Orlandini}
\affiliation{INFM-Dipartimento di Fisica, Universit\`a di Padova,
I-35131 Padova, Italy.}
\author{A.~L.~Stella}
\affiliation{INFM-Dipartimento di Fisica, Universit\`a di Padova,
I-35131 Padova, Italy.} \affiliation{Sezione INFN, Universit\`a di
Padova, I-35131 Padova, Italy.}

\date{\today}

\begin{abstract}
A lattice model of RNA denaturation which fully accounts for the excluded
volume effects among nucleotides is proposed. 
A numerical study shows that interactions forming pseudoknots must 
be included in order to get a sharp continuous transition.
Otherwise a smooth crossover occurs from the swollen linear polymer 
behavior to highly ramified, almost compact conformations with secondary
structures. 
In the latter scenario, which is appropriate when these structures are much
more stable than pseudoknot links,
 probability distributions for the lengths of both loops and main branches 
obey scaling with nonclassical exponents.
\end{abstract}

\pacs{87.15.Aa, 87.14.Gg, 05.70.Fh, 64.60.Fr}

\maketitle

In recent years, considerable attention has been devoted to the
problem of describing the formation of secondary structure 
(base pairing map) in single molecular strands of 
RNA~\cite{Zuker1989,Hofacker_FSBTS,Higgs1996,TinocoBustamante1999,BundschuhHwa1999,Pagnani_PR_2000,ChenDill2000,Higgs2000}. 
The solution of such a problem is 
regarded as an important step within the general program of understanding
how structure is encoded in the primary sequence of biopolymers.
By making use of some simplifications, like that of disregarding
excluded volume effects or pseudoknots formation,
some studies established the existence of a molten phase
at relatively high temperatures for an RNA molecule in dilute 
solution~\cite{BundschuhHwa2002, BundschuhHwa2002b, Muller2002}. 
In this phase the inhomogeneities associated
to a specific primary sequence should be irrelevant for the
large scale behavior and should allow the coexistence of a very
large number of different secondary structures of comparable free energy.

As the temperature $T$ increases, a long
RNA molecule should pass from the molten phase to a regime in which 
secondary structures essentially disappear and the global behavior
becomes that of a linear polymer chain in good solvent.
Excluded volume should
play a relevant role at such a denaturation transition. Indeed, there 
the entropic free energy gain associated to the formation of hairpins or of
more complicated branched structures with loops is
comparable with the corresponding base pair binding and staking energies, 
and depends crucially on the repulsive interactions.
Recent studies have shown that the  discontinuous nature and the universal 
features of double stranded DNA denaturation are determined by excluded volume 
interactions~\cite{caus00,kafr00, carl02}. 

To our knowledge, starting with the related pioneering 
work of de Gennes~\cite{deGennes1968}
on the statistics of branchings and hairpin helices in the 
periodic dAT copolymer, excluded volume effects were never fully
 taken into account in studies of 
RNA denaturation. This leaves open the 
problem of establishing the existence and of determining the possible character
of this transition in the long chain limit.
A realistic embedding of the system in space, taking into account excluded 
volume, is also a necessary condition for discussing 
pseudoknots and their consequences.
Pseudoknots occur, e.g.,  when two loops locally bind to each other
determining a deviation of the configuration from  planar 
topology~\cite{Higgs2000}.
Normally they are not included in models of the secondary 
structure~\cite{NussinovJacobsen1980,Zuker1989,Hofacker_FSBTS,Higgs1996,BundschuhHwa1999,Pagnani_PR_2000}, 
or are considered as a perturbation~\cite{OrlandZee2002}.

In this Letter we propose a model of the large scale
conformational behavior of RNA in the high $T$ and molten phases. 
Although schematic, our model takes fully into account excluded 
volume and allows control of the effects of pseudoknots. 
While providing useful informations on the behavior of 
finite RNA chains, an extensive numerical analysis allows to draw
precise scenarios for denaturation and the associated  scaling 
regimes in different conditions.

At coarse-grained level
we model a conformation of the RNA strand as a two-tolerant trail of  
$N$ steps on the face centered cubic (FCC) lattice~\footnote{
The FCC lattice allows for closed loops also of 
odd length and thus enriches their sampling.
}. 
This is a random walk
in which no more than two steps are allowed to overlap on a single lattice
bond, forming what we call a contact. 
This restriction takes into account the excluded volume.
In addition, by giving an orientation to the trail,
we impose that only pairs of antiparallel steps can form contacts,
and whenever this happens a gain in energy $\epsilon<0$ is counted.
The orientation of the trail reflects the backbone directionality
or RNA~\cite{Higgs2000}. In our model a contact corresponds
to a sequence of bound base pairs over a distance of the order of the 
persistence length of the RNA double helix. 
This persistence length can vary with $T$ near a denaturation transition. 
However, this effect should not matter for the large scale properties.
Since our model is coarse-grained and we are not interested in $T$'s 
below those of the molten phase, we neglect also the 
heterogeneity of base pair interactions. Thus, $\epsilon$ 
represents an effective, average parameter.

\begin{figure*}[!tbp]
\includegraphics[angle=0,width=\WIDTHAB]{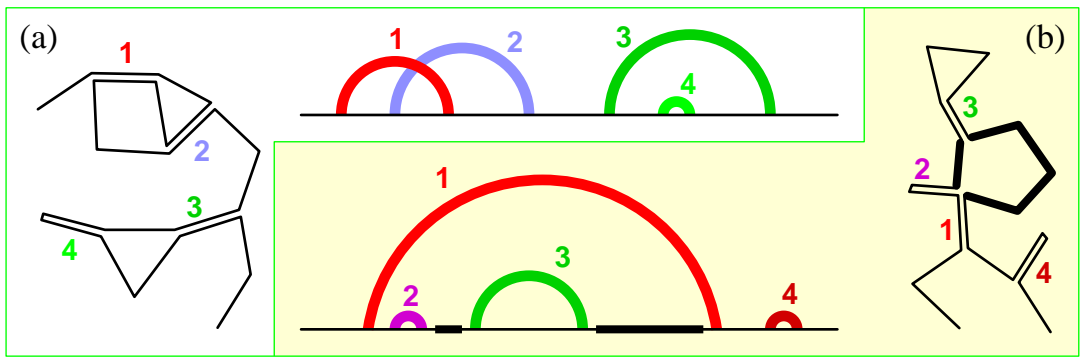}
\caption{ \label{fig:scheme}
RNA configurations and corresponding contact maps. 
Overlapped steps (contacts) are slightly split.
In (a) a pseudoknot is present (crossing of bridge ``1'' with bridge ``2'').
In (b) a loop of length $\ell=5$ is marked by a thicker line, both 
on the chain and in the contact map.
Here, ``1'' and ``4'' are main bridges.
}
\end{figure*}

Figure \ref{fig:scheme} reports schematically two possible configurations of
our model. In both (a) and (b) a
diagram on the right summarizes the corresponding contact map.
A bridge in the diagrams connects each pair of steps forming a contact.  
Bridges are numbered in order of appearance if one follows
the trail orientation. A main bridge is a bridge which is not 
inscribed within other, larger bridges.  
Unlike (b), (a) shows a pseudoknot, indicated by the crossing of two bridges
 in the diagram. This crossing means that
a step forming a loop overlaps with one outside the loop.
In order to investigate the role of pseudoknots,
we consider two variants of the model, which we refer to as~I and~II. 
While in model~I configurations with pseudoknots are 
allowed, in addition to those without pseudoknots, in model~II the former 
are forbidden altogether. The choice of
attributing the same energy to all kinds of contacts is a
simplification of model~I. Computationally it would be awkward to
attribute selectively a weaker binding energy to those 
contacts which form pseudoknots, as physically appropriate 
in most situations~\cite{TinocoBustamante1999}.

Thermodynamic quantities and canonical averages 
are defined in terms of the partition function $Z=\sum_w \exp(-H(w)/T)$. 
The sum extends to all
allowed configurations $w$ with $|w|=N$ steps, and $H=\epsilon\,N_c(w)$,
$N_c(w)$ being the number of contacts in $w$.
For both models, we sampled configurations by a multiple Markov chain
Monte Carlo procedure~\cite{Tesi_ROW96} using several ($\approx 20$)
temperatures satisfying $0 \le \epsilon/T \le 3.5$~\footnote{
To increase the mobility of the Markov chain at low $T$,
where branched structures are expected, in addition to the
usual set of local and pivot moves (see~\cite{Tesi_ROW96} and references 
therein), we implemented a variant of the pivot move, in which only one arm of
the branched structure is rotated.
In order to obtain smooth plots,
data were processed by means of the multiple histogram 
method~\cite{FerrenbergSwendsen1988}.
}.

We first computed as a function of $T$ the specific heat
of model~I and II for different $N$. At a continuous conformational transition
with crossover exponent $\phi<1$ one expects a singular behavior
$C_{\text{max}} \sim N^{2\phi -1}$ for the maximum of this
quantity as $N \to \infty$~\cite{Vanderzande,Orlandini_SS_2000}. 
In case $\phi<1/2$ such 
singularity does not imply a divergence. 
For both models we find no evidence of a diverging $C_{\text{max}}$. 
Hence, at this level we can only conclude
that for both models the denaturation transition must be continuous 
and with $\phi < 1/2$, if it exists. 

\begin{figure}[!bp]
\includegraphics[angle=-90,width=\WIDTHB]{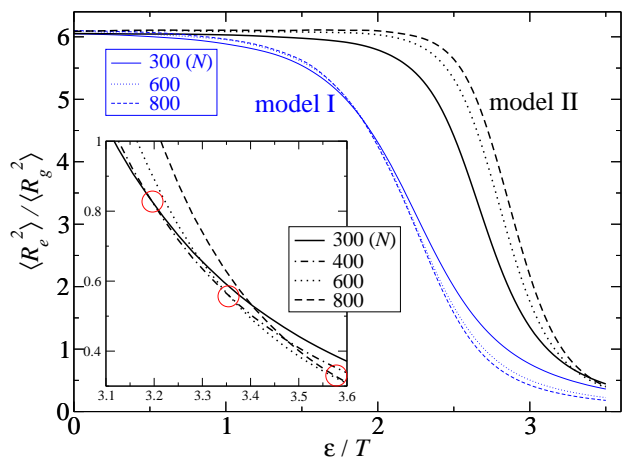}
\caption{ \label{fig:ReRg}
$\langle R_e\rangle^2/\langle R_g\rangle^2$ as a function of $\epsilon / T$ for
three different values of $N$. 
Inset: detail of the crossings of four curves for
model~II. The circles enclose intersections
between the curve pairs $(300,\,400)$, $(400,\,600)$ and $(600,\,800)$.  
}
\end{figure}

We also determined two geometrical radii of the configurations, namely
the end-to-end distance, $R_e$, and the radius of gyration 
with respect to the center of mass, $R_g$. Multicritical
phenomena theory~\cite{Privman_HA_1991,Vanderzande} 
has taught us that the ratios 
of the averages of such radii in the $N \to \infty$ limit are universal 
numbers characteristic of the different regimes involved in the transition.
Fig.~\ref{fig:ReRg} shows plots of 
$\langle R_e^2 \rangle/\langle R_g^2\rangle$ for different $N$.
For model~I the trend of the curves gives
indication of a sharp transition at $\epsilon / T \approx 1.9$.
Indeed, for high $T$ the ratio approaches from below the universal
value $6.25(1)$ appropriate for a polymer in the swollen, self avoiding walk
(SAW) regime~\cite{Li_MS_1995}. 
On the other hand, at very low $T$'s the trail should fold in double 
structures with maximal number of contacts ($N_c \sim N/2$),
in which $R_e$ necessarily approaches zero. 
This explains the trend towards zero (from above) of the curves at low $T$. 
Remarkable is the accumulation of intersection points for 
$\epsilon / T \approx 1.9$. These intersections
mark a change of the trend of the curves for increasing $N$ and suggest
the presence of a peculiar transition regime with universal 
ratio $\approx 4.8$. Hence, for model~I there is clear evidence of a
second order transition at $\epsilon / T \approx 1.9$.
For model~II there is no similar indication: the intersections are pushed
towards lower and lower $T$'s as one compares
curves corresponding to pairs of increasing $N$ values
(Fig.~\ref{fig:ReRg}, inset). This means that the larger $N$, the
deeper the SAW regime extends in the low $T$ region. 
The whole pattern suggests for model~II a smooth crossover, 
not a transition.
Further insight is provided by the study of
some scaling properties. The radius of gyration
is expected to scale as $\langle R_g\rangle \sim N^{\nu}$ for large
$N$~\cite{Vanderzande}.
For both models we observe that at high $T$ the  determinations
of $\nu$ at finite $N$,
for $N \to \infty$ approach a value $\approx 0.59$ appropriate
for a SAW in $d=3$~\cite{Vanderzande,Li_MS_1995}. 
For model~I, at $T$'s sufficiently below the transition
the $\nu$ estimates can be extrapolated to $\approx 0.35$ for large $N$. 
This indicates that the configurations are very close to compact 
in the low $T$, molten phase ($\nu \approx 1/d=1/3$). 
For model~II at very low $T$'s we extrapolate $\nu \approx 0.4$
which is also not far from $\nu=1/2$, as expected for branched 
polymers~\cite{Vanderzande},~\footnote{Two 
tolerant trails have been used to model collapse from SAW to branched polymer
behavior~\cite{Orlandini_SST_1992}. 
See also P.~Leoni and C.~Vanderzande, to be published.}.

The different behaviors of the two models are due 
to the presence of pseudoknots in model~I. 
Indeed, the fraction of sampled 
configurations with pseudoknots in model~I is already substantial 
and increases with $N$ at high $T$. 
It reaches soon values close to $1$ near the transition and below.  
Pseudoknots correspond to the
formation of extra binding contacts and thus can lead to
more compact configurations with respect to the case of model~II. 
These extra contacts trigger the sharp transition observed at 
$\epsilon / T\approx 1.9$. 
For model~II, if present, a transition should be located at much lower
$T$'s, most likely below the range of applicability 
of the model~\footnote{
At low enough $T$ the molten phase is supposed to leave place
to a glassy one, in which the inhomogeneities due to primary structure
become relevant~\cite{Higgs1996,BundschuhHwa1999,BundschuhHwa2002,BundschuhHwa2002b,Pagnani_PR_2000}.}. 

The possibility of forming pseudoknots is a
driving factor in tertiary structure formation~\cite{TinocoBustamante1999}. 
However, model~I somehow overamplifies this factor, because it gives pseudoknot
forming contacts an energy equal to that of the other contacts. 
In fact contacts forming pseudoknots should correspond most often to
the weak binding of quite short portions of single strand loops, like
for kissing hairpins. 
Longer bindings giving rise to pseudoknots are expected to be kinetically 
inhibited~\cite{BundschuhHwa1999}. One way to make the energies of 
contacts forming pseudoknots closer to those of the other contacts
is to introduce sufficiently high concentrations of divalent metal ions, 
like $\text{Mg}^{2+}$
in solution~\cite{MisraDraper1998,TinocoBustamante1999}.
On the other hand, in model~II pseudoknot forming contacts would appear 
if one would look at the details of the configurations at somewhat more
coarse-grained level. This means that this model can be interpreted
as one similar to model~I, but giving essentially zero
energy to such contacts. Thus,
it is reasonable to expect the description of model~II to be
most appropriate for not too low $T$'s, and expecially when, e.g.,
a low concentration of $\text{Mg}^{2+}$ ions in solution
enlarges the stability gap between secondary structure and pseudoknot
forming links~\cite{MisraDraper1998,TinocoBustamante1999}.

\begin{figure}[!tbp]
\includegraphics[angle=-90,width=\WIDTHB]{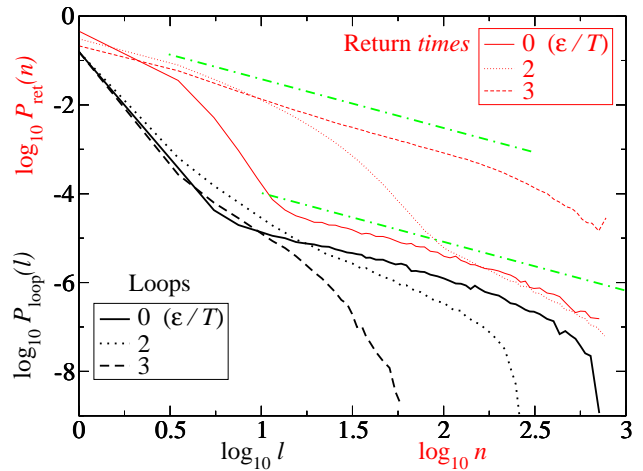}
\caption{ \label{fig:PlPn}
Log-log plots for {$P_{\text{loop}}(\ell)$} 
(thick lines, shifted down by 0.5 for clarity) 
and {$P_{\text{ret}}(n)$} (thin lines, red online), 
both for $N=800$ and for different $T$ values.
The dot-dashed lines have slope $-1.1$. 
}
\end{figure}

RNA denaturation  corresponds to a substantial suppression
with increasing $T$ of the highly ramified structure of loops and branches 
characterizing the molten phase. The analysis of the loops is  feasible and 
particularly instructive in model~II.  
For DNA, the distribution of the lengths of
denaturated loops, corresponding to openings of the double helix,
follows a power law whose exponent $c$ determines 
the character of the transition~\cite{kafr00,carl02}. 
A simple example of loop in RNA is given by the closure of an isolated 
hairpin. In this case the loop is connected 
to the rest of the structure by a single branch of double steps in model~II. 
Of course, more complicated situations may occur 
(thick loop in Fig.~\ref{fig:scheme}(b)). 
Even at high $T$ an extensive number of minor spike-like branches is present 
along the RNA backbone. 
Thus, loops with a fixed number of branches naturally have 
length distributions with rather short cut-off.
Therefore, we decided to sample the length of
all loops identified in the various configurations, irrespective
of the number of outgoing double step branches. 
The various lengths can be obtained from the contact map
of each configuration, by using a recursive algorithm that identifies 
all the loops inside each main bridge in the diagram.
Another interesting quantity is the return {\it time},
i.e.\ the total number of steps comprised within a main
bridge. In the assumed planar topology of model~II this time is
the total arc length corresponding to each departure of the configuration from
a contact-free, linear polymer behavior.
Probability distributions of the loop lengths, $\ell$,
and of the return times, $n$, are plotted in Fig.~\ref{fig:PlPn}
for different $T$'s and for $N=800$.
At high $T$, after transients both distributions behave as 
power laws with approximately identical exponents: 
$P_{\text{loop}}(\ell)\sim \ell^{-c_\ell}$, 
$P_{\text{ret}}(n)\sim n^{-c_r}$, with
$c_\ell \simeq c_r = 1.1(1).$
The peaks at small $\ell$ and $n$ in the distributions
indicate that loops at high $T$ mostly occur within isolated
small hairpins in model~II. 
The identity of $c_\ell$ and $c_r$ 
means that almost all large bridges are also main bridges.
Thus, the return time essentially coincides with the loop length at high $T$.
At lower $T$, while $P_{\text{loop}}$ becomes
shorter and shorter ranged for decreasing $T$, the behavior of 
$P_{\text{ret}}$ remains of power law type at large arguments. The value of 
$c_r$ remains stable and close to that estimated for $\epsilon / T=0$.
This means that as the RNA molecule enters deeper and
deeper into the molten phase with developed secondary structures,
the loops become shorter and shorter. 
On the other hand,  the main branches departing from the 
contact-free backbone encompass all accessible length scales, as 
appropriate for a branched polymer. 
This could also explain why in this range of $T$ the exponent $\nu$ 
discussed above is not far from $1/2$, as for branched 
polymers~\cite{Vanderzande}.
The exponent $c_r$ obtained here definitely deviates from 
the mean field value $3/2$~\cite{BundschuhHwa2002}.
 
Summarizing, for model~I we could establish the existence of a sharp
denaturation transition which is due to the presence of
pseudoknots. Unlike the melting of DNA, this is a second order transition.
Model~I applies to experimental situations in which the stability
gap between secondary and tertiary folding is sensibly reduced, e.g.,
by a high concentration
of  $\text{Mg}^{2+}$ ions~\cite{MisraDraper1998,TinocoBustamante1999}.
On the other hand, one
can regard model~II as a more adequate description of
RNA when the same stability gap  is large,
e.g. with low $\text{Mg}^{2+}$ 
concentrations~\cite{MisraDraper1998,TinocoBustamante1999}. 
In this model there is no sharp transition and the denaturation 
occurs as a crossover from linear to branched-compact polymer behavior. 
The geometry of this crossover is well described by
the distributions $P_{\text{loop}}$ and $P_{\text{ret}}$ and by
their exponents, whose nonclassical values are a 
further consequence of excluded volume. 

We acknowledge discussions with E.~Carlon. The work was supported by 
MIUR-COFIN01 and by INFM-PAIS02.

\end{document}